\documentstyle[12pt,a4,epsfig]{article}
\begin{document}
\vspace{2cm}
\begin{center}
{\Large Determination of the strangeness contents of light-flavour isoscalars 
from its production rates in hadronic Z decays measured at LEP}

\bigskip
\bigskip
{\large Vladimir UVAROV}

Institute for High Energy Physics, RU-142284, Protvino, Russia
\end{center}

\bigskip
\bigskip
\noindent
\small
{\bf Abstract}

\smallskip
A new phenomenological approach is suggested to determine the strangeness 
contents of light-flavour isoscalars.
This approach is based on phenomenological laws of hadron production related 
to the spin, isospin, strangeness content and mass of the particles.
The ``effective'' numbers of $s$ and $\bar{s}$ quarks in the isoscalar 
partners $i_1$ and $i_2$ are given by the nonstrange-strange mixing angle 
$\varphi$: $k(i_1)=2\sin^2\varphi$ and $k(i_2)=2\cos^2\varphi$. 
From the total production rates per hadronic Z decay of all light-flavour 
hadrons measured so far at LEP the values for $k$ are found to be:
$k(\eta)$ $\equiv$ 2$-$$k(\eta^{\,\prime})$ = 0.91$\pm$0.12, 
$k(\phi)$ $\equiv$ 2$-$$k(\omega)$ = 1.94$\pm$0.09, 
$k(f_2^{\prime})$ $\equiv$ 2$-$$k(f_2^{~})$ = 1.84$\pm$0.21 and 
$k(f_0)$ = 0.09$\pm$0.13.
Our results on the $\eta$--$\eta^{\,\prime}$, $\omega$--$\phi$ and 
$f_2^{~}$--$f_2^{\prime}$ isoscalar mixing are consistent with the present
experimental evidence. 
Quite remarkably, our value for $k(\eta)$ corresponds to the singlet-octet 
mixing angle $\theta_P$ = $-$12.4$^{\circ}$$\pm$3.5$^{\circ}$.
The obtained strangeness content of the $f_0$(980) scalar/isoscalar
is not consistent with the values supported by different model studies.
However, taking the $f_0$(980) in our analysis with the mass of the bare state 
($K$-matrix pole) $f_0^{bare}$(720$\pm$100), the mixing angle is found to be:
$\vert\varphi_S^{\,bare}\vert$ = 73$^{\circ}$$\pm$7$^{\circ}$$\pm$24$^{\circ}$,
in good agreement with the prediction of the $K$-matrix analysis.
\normalsize

\bigskip
\bigskip
The quark contents of pseudoscalar ($P$), vector ($V$), tensor ($T$)
and scalar ($S$) mesons have been discussed many times from the beginning 
of the creation of unitary SU(3)-flavour symmetry. 
This is a quite interesting question because the quark contents of the 
lightest isoscalars are different from the prediction of the SU(3) quark 
model.
The phenomenological studies of hadronic processes involving isoscalars 
usually make assumptions about their quark compositions. 
Therefore, to the understanding of the quark model and QCD, it is very 
important to determine the SU(3)-breaking hadronic parameters which define 
these quark compositions. 

In terms of the SU(3) singlet and octet basis states
\begin{equation}
\label{eq1}
\eta_{1} = (u\bar{u}+d\bar{d}+s\bar{s})/\sqrt{3}, ~~~~~
\eta_{8} = (u\bar{u}+d\bar{d}-2s\bar{s})/\sqrt{6}
\end{equation}
the quark contents of the physical $\eta$ and $\eta^{\,\prime}$ states
are given by the singlet-octet mixing angle $\theta_P$.
Assuming the orthogonality of the physical states and no mixing with other 
isoscalars and glueballs the flavour wave functions of the $\eta$ and 
$\eta^{\,\prime}$ pseudoscalars are defined to be:
\begin{equation}
\label{eq2}
\eta~ = \eta_8\cdot\cos\theta_P - \eta_1\cdot\sin\theta_P, 
\end{equation}
\begin{equation}
\label{eq3}
\eta^{\,\prime} = \eta_8\cdot\sin\theta_P + \eta_1\cdot\cos\theta_P.
\end{equation}
To determine the {\it strangeness} contents of the same isoscalars it is 
more convenient to use the so-called nonstrange-strange quark basis
($n\bar{n} = (u\bar{u}+d\bar{d})/\sqrt{2}$ and $s\bar{s}$):
\begin{equation}
\label{eq4}
\eta~ = n\bar{n}\cdot\cos\varphi_P - s\bar{s}\cdot\sin\varphi_P, 
\end{equation}
\begin{equation}
\label{eq5}
\eta^{\,\prime} = n\bar{n}\cdot\sin\varphi_P + s\bar{s}\cdot\cos\varphi_P,
\end{equation}
where $\varphi_P$ is the nonstrange-strange mixing angle (with 
$\theta_P = \varphi_P - \arctan\sqrt{2}$\,).
The quark contents of the $\omega$--$\phi$ and $f_2^{~}$--$f_2^{\prime}$ 
isoscalars are defined in a way analogous to the 
$\eta$--$\eta^{\,\prime}$ case, replacing 
$\eta \rightarrow \phi (f_2^{\prime})$,
$\eta^{\,\prime} \rightarrow \omega (f_2^{~})$ 
in Eqs.~(\ref{eq2})--(\ref{eq3}) and
$\eta \rightarrow \omega (f_2^{~})$,
$\eta^{\,\prime} \rightarrow \phi (f_2^{\prime})$ 
in Eqs.~(\ref{eq4})--(\ref{eq5}).
The flavour wave function of the $f_0$(980) scalar meson is written as
\begin{equation}
\label{eq6}
f_0 = n\bar{n}\cdot\cos\varphi_S + s\bar{s}\cdot\sin\varphi_S.
\end{equation}

The values of the mixing angles have been estimated from different  
phenomenological and theoretical analyses (see Refs. \cite{I1}-\cite{I9} and 
references therein).
The phenomenological estimations of the pseudoscalar mixing angle 
$\theta_P$ (or $\varphi_P$)
use the available world data on the following decay processes:
strong decays of tensors and higher-spin mesons ($M_{J>2}$) into pseudoscalar 
pairs,
radiative transitions between vectors and pseudoscalars,
two-photon annihilation decays,
leptonic decays of vectors,  
$J/\psi$ decays into a vector plus a pseudoscalar, 
radiative $J/\psi$ decays,  
semileptonic $D_s$ decays plus transition form factors
$\eta/\eta^{\,\prime} \rightarrow \gamma\gamma^*$. 
There are also phenomenological analyses of $\theta_P$ which use the data 
on the quasi-two-body $\pi^-p$ and $\bar{p}p$ reactions.  
The values for $\theta_P$ (and $\varphi_P$) obtained from the above highlighted 
processes are given in Table~\ref{tab1} together with the corresponding 
references.
\begin{table}[ht]
\small
\caption{\small Compilation of the values of the $\eta$--$\eta^{\,\prime}$ 
mixing angles $\varphi_P$ and $\theta_P$ obtained from different processes 
(with $\theta_P = \varphi_P - \arctan\sqrt{2} \simeq 
\varphi_P - 54.7^{\circ}$).}
\begin{center} 
\begin{tabular}{|l|c|c|c|}
\hline
process&$\varphi_P$\,($\,^{\circ}$)&$\theta_P$\,($\,^{\circ}$)&Ref.\\
\hline
\hline
$T(2^{++}) \rightarrow PP$&$42\pm 2$&$-13\pm 2$&\cite{R1}\\
~&$43.1\pm 3.0$&$-11.6\pm 3.0$&\cite{R10}\\
\hline
$M_{J>2} \rightarrow PP$&$41\pm 4$&$-14\pm 4$&\cite{R1}\\
\hline
$V \rightarrow P\gamma$;~ $P \rightarrow V\gamma$&
$37.7\pm 2.4$&$-17.0\pm 2.4$&\cite{R2}\\
~&$35.3\pm 5.5$&$-19.4\pm 5.5$&\cite{R10}\\
\hline
$P \rightarrow \gamma\gamma$&$41.3\pm 1.3$&$-13.4\pm 1.3$&\cite{R1}\\
~&$36.3\pm 2.0$&$-18.4\pm 2.0$&\cite{R3}\\
\hline
$P \rightarrow V\gamma / \gamma\gamma$;~ $V \rightarrow P\gamma / e^+e^-$& 
$43.1\pm 0.8$&$-11.6\pm 0.8$&\cite{R12}\\
\hline
$J/\psi \rightarrow VP$&$35.5\pm 1.4$&$-19.2\pm 1.4$&\cite{R4}\\
~&$35.6\pm 1.4$&$-19.1\pm 1.4$&\cite{R5}\\
~&$37.8\pm 1.7$&$-16.9\pm 1.7$&\cite{R1}\\
~&$39.9\pm 2.9$&$-14.8\pm 2.9$&\cite{R10}\\
\hline
$J/\psi \rightarrow P\gamma$&$39.0\pm 1.6$&$-15.7\pm 1.6$&\cite{R10}\\
\hline
$D_s \rightarrow P\,e\,\nu_{\,e}$;~
$\eta/\eta^{\,\prime\,} \rightarrow \gamma\gamma^*$&
$38.0\pm 2.8$&$-16.7\pm 2.8$&\cite{R9}\\
~&$41.2\pm 4.7$&$-13.5\pm 4.7$&\cite{R11}\\
\hline
$\pi^-p \rightarrow (\eta/\eta^{\,\prime\,})\,n$&
$36.5\pm 1.4$&$-18.2\pm 1.4$&\cite{R6}\\
~&$39.3\pm 1.2$&$-15.4\pm 1.2$&\cite{R7}\\
\hline
$\bar{p}p \rightarrow (\eta/\eta^{\,\prime\,})\,(\pi^0/\eta/\omega)$&
$37.4\pm 1.8$&$-17.3\pm 1.8$&\cite{R8}\\
\hline
{\normalsize \it average}&$39.2\pm 1.3$&$-15.5\pm 1.3$&\cite{R1,R10}\\
\hline
\hline
${\rm Z}^0 \rightarrow hadrons$&
$42.3\pm 3.5$&$-12.4\pm 3.5$&this analysis\\
\hline
\end{tabular}
\label{tab1}
\end{center}
\normalsize
\end{table}
The theoretical analyses of the mixing angles are usually model dependent, and 
the predictions for $\theta_P$ range from $-23^{\circ}$ to $-10^{\circ}$.
Classic examples are the quadratic and linear Gell-Mann--Okubo (GMO) mass 
formulae which yield the values $\theta_P^{\,quad} \simeq -10^{\circ}$
and $\theta_P^{\,lin} \simeq -23^{\circ}$ \cite{PDG}.
For the vector ($\varphi_V$) and tensor ($\varphi_T$) mixing angles the most of 
theoretical and phenomenological analyses (see, for example, Refs.
\cite{PDG,R1,R2,R12,R13,R14,R15}) predict the values which are very close to 
the ``ideal'' mixing: 
$\varphi_V = \theta_V - \theta^{\,ideal} \simeq +3.4^{\circ}$ and 
$\varphi_T = \theta_T - \theta^{\,ideal} \simeq -7.3^{\circ}$
($\theta^{\,ideal} = \arctan{1/\sqrt{2}} \simeq 35.3^{\circ}$).

The interpretation of the $f_0$(980) scalar meson is one of the most 
controversial in meson spectroscopy \cite{I10}. 
The question is whether the $f_0$(980) consists mostly of nonstrange or of 
strange quarks.
The recent phenomenological analyses of the experimental data on the 
following decay processes $\phi \rightarrow \pi^0\pi^0\gamma$, 
$J/\psi \rightarrow \omega\pi\pi$ \cite{I11}, 
$D_s^+ \rightarrow f_0(980)\pi^+$ \cite{I9} and
$f_0(980) \rightarrow \pi\pi/K\bar{K}/\gamma\gamma$ \cite{I8} favour
the $s\bar{s}$ dominance of the $f_0$(980). 
The phenomenological study of low-energy $\pi\pi$ scattering \cite{I12}
and the phase shift analysis of $\pi K$ scattering \cite{I13} have led to the 
same conclusion.

In the present paper the strangeness contents of light-flavour isoscalars are 
obtained for the first time from its total production rates per hadronic Z 
decay measured at LEP. 
The new decay process (${\rm Z}^0 \rightarrow hadrons$) is added to the 
above-listed processes used to determine 
the mixing angles.  

Recently it has been shown \cite{U1}-\cite{U4} that the total production rates 
per hadronic Z decay ($\langle n \rangle$) of all light-flavour mesons ($M$) 
and baryons ($B$) measured so far at LEP follow phenomenological laws related 
to the spin ($J$), isospin ($I$), strangeness content and mass ($m$) of 
the particles. These regularities can be combined into one empirical formula:
\begin{equation}
\label{eq7}
{\langle n \rangle} \,=\, 
A\cdot\beta_H\cdot(2J+1)\cdot\gamma^{\,k}\cdot\exp{[-b_H(m/m_0)^{N_H}]},
\end{equation}
where $H$ = $M$ or $B$,\, $m_0=1$ GeV/$c^2$,\, $\gamma$ is the strangeness 
suppression factor with a value of $\gamma \simeq 0.5$ for all hadrons,
$k$ is the number of $s$ and $\bar{s}$ quarks in the hadron and $b_H$
is the slope of the mass dependence.
The values of the degree $N_H$ and of the coefficient $\beta_H$ are different 
for mesons and baryons:
\begin{equation}
\label{eq8}
N_M = 1,~~ N_B = 2 ~~~{\rm and}~~~
\beta_M = 1,~~ \beta_B = {4\over{C_{\pi/p}\cdot\lambda_{\,QS}}}, 
\end{equation}
where $\lambda_{QS} = (2J+1)(2I+1)$ can be interpreted as a fermion 
suppression factor originating from quantum statistics properties of bosons 
and fermions and $C_{\pi/p}$ is the $\pi$/p ratio at the zero mass limit 
with a value of $C_{\pi/p} \simeq 3$ which could be expected from quark
combinatorics. 

According to the results of Refs.~\cite{U2,U3,U4} the normalization parameter 
$A$ in Eq.~(\ref{eq7}) is the same for all mesons, but the meson slope $b_M$ 
is split into two: one for vector, tensor and scalar mesons ($b_{V,T,S}$) 
and another for pseudoscalar mesons ($b_P$). 
The slope splitting of the mass dependence of meson production rates can 
probably be explained by the influence of the spin-spin interaction between 
the quarks of the meson (the spins of quarks are parallel for vector, tensor 
and scalar mesons and anti-parallel for pseudoscalar mesons). However, there 
is no influence of the value and orientation (with respect to the net spin) 
of the orbital angular momentum of the quarks, i.e. of the spin-orbital 
interaction of the quarks. 

In the analyses \cite{U1}-\cite{U4} the strangeness contents $k$ of baryons 
and $I$$\neq$0 mesons were taken from the prediction of the SU(3) quark model.
For isoscalars the following values for $k$ were used:
$k(\omega)=k(f_2^{~})=k(f_0)=0$, $k(\eta)=k(\eta^{\,\prime})=1$ and
$k(\phi)=k(f_2^{\prime})=2$.
The {\it purpose of the present paper} is to obtain 
the values of the ``effective'' numbers $k$ of the $\eta$--$\eta^{\,\prime}$, 
$\omega$--$\phi$, $f_2^{~}$--$f_2^{\prime}$ and $f_0$(980) isoscalars 
from the fit of Eq.~(\ref{eq7}) to the total production rates per hadronic Z 
decay of all light-flavour hadrons measured so far at LEP 
and then to determine the corresponding mixing angles.
The {\it basic idea} of this analysis is the following relations between the 
numbers $k$ and the nonstrange-strange mixing angles $\varphi$:
\begin{equation}
\label{eq9}
k(\eta)\, \,\equiv\, 2 - k(\eta^{\,\prime}) \,=\, 2 \sin^2\varphi_P,  
\end{equation}
\begin{equation}
\label{eq10}
k(\phi)\, \,\equiv\, 2 - k(\omega)\,\, \,=\, 2 \cos^2\varphi_V,  
\end{equation}
\begin{equation}
\label{eq11}
k(f_2^{\prime}) \,\equiv\, 2 - k(f_2^{~}) \,=\, 2 \cos^2\varphi_T,  
\end{equation}
\begin{equation}
\label{eq12}
k(f_0) \,=\, 2 \sin^2\varphi_S,  
\end{equation}
which assume the orthogonality of the physical isoscalar partners and 
no mixing with other isoscalars and glueballs.

The total\,\footnote{~The quoted rates include decay products from resonances 
and particles with $c\tau < 10$ cm.} production rates per hadronic Z decay
of light-flavour hadrons, used in this analysis, were obtained for at least one 
state of a given isomultiplet as a weighted-average\,\footnote{~In calculating 
the errors of averages, the standard weighted least-squares procedure 
suggested by the PDG~\cite{PDG} was applied: if the quantity 
$[\chi^2/(N-1)]^{1/2}$ was greater than 1, the error of the average was 
multiplied by this scale factor.} of the measurements of the four LEP 
experiments: ALEPH~\cite{A1}-\cite{A3}, DELPHI~\cite{D9},\cite{D1}-\cite{D8}, 
L3~\cite{L1}-\cite{L4} and OPAL~\cite{O1}-\cite{O9}. 
Then, in the fits of our analysis, we use the {\it production rates per 
isospin state} ($\langle n \rangle$) which were obtained by averaging the 
above-obtained production rates of particles belonging to the same 
isomultiplet (${\langle n \rangle}_i$):
\begin{equation}
\label{eq13}
{\langle n \rangle} \,=\, 
{1\over X}\,\sum_{i\,=\,1}^{i\,=\,X}{\langle n \rangle}_i,
\end{equation}
where $X \leq (2I+1)$ is the number of {\it measured} isospin states. 
The exact definitions (\ref{eq13}) and the experimental values 
of the data points $\langle n \rangle$ are given in Table~\ref{tab2} 
together with the corresponding references.
\begin{table}[htbp]
\small
\caption{\small Definitions and values of the total production rates per 
isospin state of light-flavour hadrons in hadronic Z decays obtained from 
the weighted-average values of the four LEP experiments. The baryon rates 
include charge conjugated states.}
\begin{center}
\begin{tabular}{|l|c|c|l|}
\hline
hadron&definition&$\langle n \rangle$&References\\
\hline
\hline
$\pi$&{$1\over3$}\,($\pi^0$+$\pi^{\pm}$)&
8.83$\pm$0.15&\cite{A1,A2,D4,D7,L1,O2,O9}\\
K&{$1\over4$}\,(K$^0$+$\bar{\rm K}^0$+K$^{\pm}$)&
1.075$\pm$0.016&\cite{A2,A3,D2,D7,L3,O2,O4}\\
$\eta$&&
0.94$\pm$0.08&\cite{L1,O9}\\
$\eta^{\prime}$&&
0.17$\pm$0.05&\cite{L2,O9}\\
\hline
$\rho$&{$1\over3$}\,($\rho^0$+$\rho^{\pm}$)&
1.21$\pm$0.15&\cite{A1,D8,O9}\\
K$^*$&{$1\over4$}\,(K$^{*0}$+$\bar{\rm K}^{*0}$+K$^{*\pm}$)&
0.367$\pm$0.014&\cite{A1,D2,D6,O1,O5}\\
$\omega$&&
1.084$\pm$0.086&\cite{A1,L2,O9}\\
$\phi$&&
0.0966$\pm$0.0073&\cite{A1,D6,O8}\\
\hline
N&p&
1.037$\pm$0.040&\cite{A2,D7,O2}\\
$\Lambda$&&
0.388$\pm$0.011&\cite{A3,D3,L3,O6}\\
$\Sigma$&{$1\over3$}\,($\Sigma^-$+$\Sigma^0$+$\Sigma^+$)&
0.089$\pm$0.005&\cite{A1,D5,D9,L4,O7}\\
$\Xi$&$\Xi^-$&
0.0265$\pm$0.0011&\cite{A1,D3,O6}\\
\hline
$\Delta$&$\Delta^{++}$&
0.088$\pm$0.035&\cite{D1,O3}\\
$\Sigma^*$&{$1\over2$}\,$\Sigma^{*\pm}$&
0.0234$\pm$0.0022&\cite{A1,D3,O6}\\
$\Xi^*$&$\Xi^{*0}$&
0.0058$\pm$0.0010&\cite{A1,D3,O6}\\
$\Omega^-$&&
0.0013~$\pm$0.00024&\cite{A1,D5,O6}\\
\hline
$a_0$(980)&{$1\over2$}\,$a_0^{\pm}$&
0.135$\pm$0.055&\cite{O9}\\
$f_0$(980)&&
0.147$\pm$0.011&\cite{D8,O8}\\
\hline
K$_2^*$(1430)&{$1\over2$}\,(K$_2^{*0}$+$\bar{\rm K}_2^{*0}$)&
0.042$\pm$0.020&\cite{D8,O5}\\
$f_2$(1270)&&
0.169$\pm$0.025&\cite{D8,O8}\\
$f_2^{\prime}$(1525)&&
0.012$\pm$0.006&\cite{D8}\\
\hline
$\Lambda$(1520)&&
0.0225$\pm$0.0028&\cite{D9,O6}\\
\hline
\end{tabular}
\label{tab2}
\end{center}
\normalsize
\end{table}

\begin{figure}[hbtp]
\centering\mbox{\epsfig{file=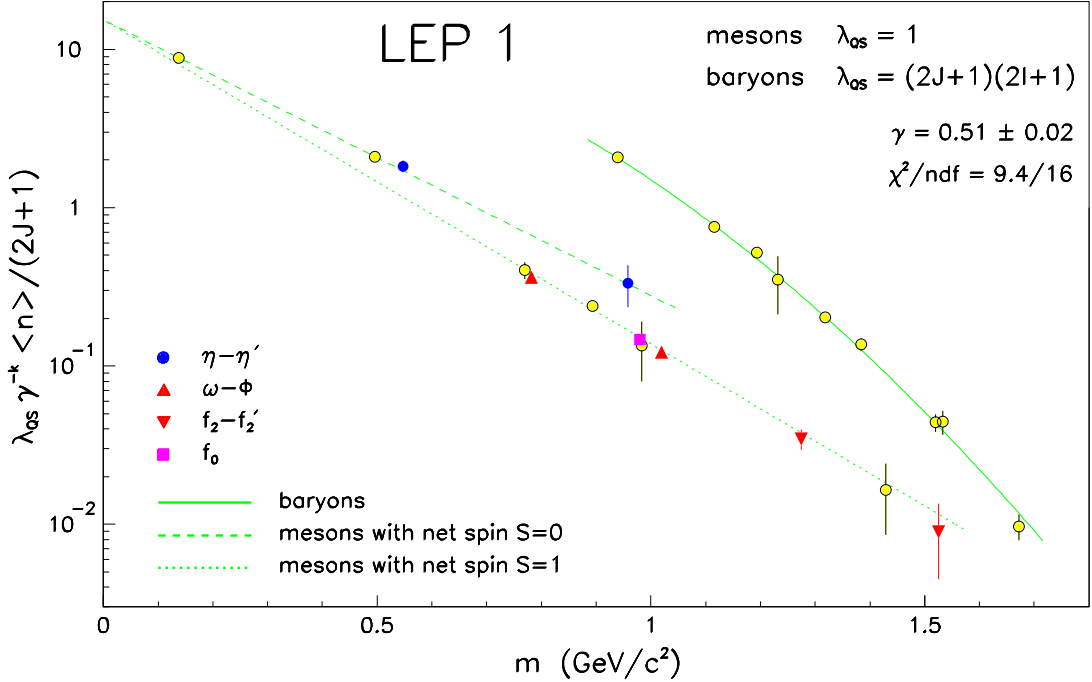,width=\textwidth}}
\caption{\small Total production rate per spin and isospin state weighted by 
a factor $\lambda_{QS}\,\gamma^{-k}$ as a function of the mass for mesons and 
baryons in hadronic Z decays. Curves are the result of the fit~1 (see text).}
\label{fig1}
\end{figure}

In comparison with the analyses of Refs.~\cite{U1}-\cite{U4}, in the present
analysis Eq.~(\ref{eq7}) is simultaneously fitted to all experimental data 
points given in Table~\ref{tab2} and it assumes the validity of the relation 
$V/P = (2J+1) = 3$ for the vector-to-pseudoscalar ratio at the zero mass limit.
In the first fit (fit~1) the numbers $k$ for the $\eta$--$\eta^{\,\prime}$, 
$\omega$--$\phi$ and $f_2^{~}$--$f_2^{\prime}$ isoscalars are fixed and given 
by Eqs.~(\ref{eq9})-(\ref{eq11}) with the values of the mixing angles predicted
by the quadratic GMO mass formula: $\varphi_P \simeq 44.7^{\circ}$, 
$\varphi_V \simeq 3.7^{\circ}$ and $\varphi_T \simeq -7.3^{\circ}$ \cite{PDG}.
The fixed value of $k=0$ is used for the $f_0$(980) scalar. 
The result of the fit~1 is shown in Fig.~\ref{fig1} and in Table~\ref{tab3}.
Fig.~\ref{fig1} shows the mass dependence of the total production rates per 
spin and isospin state for hadrons in hadronic Z decays weighted by a factor 
$\lambda_{QS}\,\gamma^{-k}$ where $\lambda_{QS}=1$ for mesons and  
$\lambda_{QS}=(2J+1)(2I+1)$ for baryons. 
Three curves on this figure are the result of the fit~1 for baryons, for 
mesons with the net spin S\,=\,0 (pseudoscalars) and for mesons with the 
net spin S\,=\,1 (vectors, tensors and scalars).
The values for $\gamma$ and $C_{\pi/p}$ are found to be 
$\gamma = 0.51\pm0.02$ and $C_{\pi/p} = 2.8\pm0.2$. 
The latter is consistent with the quark combinatorics prediction $\pi$/p = 3
for the direct production rates.
This coincidence can probably be explained by the absence of decay 
processes at zero masses.
\begin{table}[htbp]
\small
\caption{\small Values of the parameters in Eq.~(\ref{eq7}) obtained in the 
four fits of this analysis.}
\begin{center} 
\begin{tabular}{|l|c|c|c|c|}
\hline
~ & fit~1 & fit~2 & fit~3 & fit~4 \\
\hline
$A$        &15.4$\pm$0.4 &14.7$\pm$0.9 &15.4$\pm$0.4 &15.2$\pm$0.4 \\
$C_{\pi/p}$& 2.8$\pm$0.2 &~3~~~~\,fixed&~3~~~~\,fixed& 2.8$\pm$0.2 \\
$\gamma$   &0.51$\pm$0.02&0.51$\pm$0.02&0.50$\pm$0.02&0.50$\pm$0.02\\
$b_P$      &4.01$\pm$0.11&4.03$\pm$0.11&3.97$\pm$0.12&3.94$\pm$0.14\\
$b_{V,T,S}$&4.71$\pm$0.06&4.67$\pm$0.07&4.69$\pm$0.06&4.67$\pm$0.08\\
$b_B$      &2.70$\pm$0.07&2.57$\pm$0.07&2.62$\pm$0.04&2.66$\pm$0.08\\
$N_M$      &~1~~~~\,fixed&1.04$\pm$0.06&~1~~~~\,fixed&~1~~~~\,fixed\\
$N_B$      &~2~~~~\,fixed&2.07$\pm$0.05&~2~~~~\,fixed&~2~~~~\,fixed\\
$k(\eta)$  &~0.99~fixed  &~0.99~fixed  &0.90$\pm$0.12&0.91$\pm$0.12\\
$k(\phi)$  &~1.99~fixed  &~1.99~fixed  &1.94$\pm$0.09&1.94$\pm$0.09\\
$k(f_2^{\prime})$
           &~1.97~fixed  &~1.97~fixed  &1.86$\pm$0.20&1.84$\pm$0.21\\
$k(f_0)$   &~0~~~~\,fixed&~0~~~~\,fixed&~0~~~~\,fixed&0.09$\pm$0.13\\
$\chi^2/ndf$
           &9.4\,/\,16   &8.6\,/\,15   &9.8\,/\,14   &8.1\,/\,12   \\
\hline
\end{tabular}
\label{tab3}
\end{center}
\normalsize
\end{table}

In the second fit (fit~2) of our analysis we test the sensitivity of the 
empirical formula (\ref{eq7}) to the values of the degree $N_H$. This was 
not discussed in the previous analyses \cite{U1}-\cite{U4}. In the fit~2 
the numbers $k$ for the isoscalars are still fixed with the same values as 
in the fit~1, the ratio $C_{\pi/p}$ is fixed with a value of $C_{\pi/p}=3$ 
and the degrees $N_M$ and $N_B$ are free parameters. 
The values for $N_M$ and $N_B$ are found to be (see Table~\ref{tab3}):
\begin{equation}
\label{eq14}
N_M \,=\, 1.04\pm0.06 ~~~{\rm and}~~~ N_B \,=\, 2.07\pm0.05. 
\end{equation}
These values are close within the relatively small errors to the values of 1 
and 2 used previously. So, we can conclude that this test (fit~2) strongly 
suggests the use of $N_M=1$ and $N_B=2$ in our phenomenological analysis. 
These values are fixed in the next fits.

In the third fit (fit~3) the strangeness contents $k(\eta)$, $k(\phi)$
and $k(f_2^{\prime})$ are free parameters. 
The values for $k(\eta^{\,\prime})$, $k(\omega)$ and $k(f_2^{~})$
are given by the constraints (\ref{eq9})-(\ref{eq11}).
The fixed parameters in the fit~3 are: $C_{\pi/p}=3$, $N_M=1$, $N_B=2$
and $k(f_0)=0$.
The values for $k(\eta)$, $k(\phi)$ and $k(f_2^{\prime})$ are found to be 
(see Table~\ref{tab3}): $k(\eta)=0.90\pm0.12$, $k(\phi)=1.94\pm0.09$ and 
$k(f_2^{\prime})=1.86\pm0.20$.
These values agree within the errors with theoretical and phenomenological
analyses. Only one question in our analysis is still open.
The strangeness content of the $f_0$(980) scalar is fixed with
the ad hoc value of $k(f_0)=0$ which was suggested in 
Refs. \cite{U1,U2,U3,U4}.
However, the recent phenomenological analyses of different physical processes
suggest the $s\bar{s}$ dominance of the $f_0$(980). 

In our final fit (fit~4) only the values of $N_M=1$ and $N_B=2$ are fixed.
All other parameters are free.
Our final values for $\gamma$ and $C_{\pi/p}$ (see Table~\ref{tab3}) are: 
\begin{equation}
\label{eq15}
\gamma \,=\, 0.50\pm0.02 ~~~{\rm and}~~~ C_{\pi/p} = 2.8\pm0.2, 
\end{equation}
in good agreement with our previous results \cite{U2,U3,U4}.
The strangeness contents of the isoscalars, the main purpose of this paper, 
are found to be:
\begin{equation}
\label{eq16}
k(\eta)\, \,\equiv\, 2 - k(\eta^{\,\prime}) \,=\, 0.91 \pm 0.12,  
\end{equation}
\begin{equation}
\label{eq17}
k(\phi)\, \,\equiv\, 2 - k(\omega)\, \,=\, 1.94 \pm 0.09,  
\end{equation}
\begin{equation}
\label{eq18}
k(f_2^{\prime}) \,\equiv\, 2 - k(f_2^{~}) \,=\, 1.84 \pm 0.21,  
\end{equation}
\begin{equation}
\label{eq19}
k(f_0) \,=\, 0.09 \pm 0.13.  
\end{equation}
The corresponding values of the mixing angles are given by Eqs. 
(\ref{eq9})-(\ref{eq12}) with the values (\ref{eq16})-(\ref{eq19}) 
for the numbers $k$.

Our value (\ref{eq16}) for the number $k(\eta)$ corresponds to the following 
values of the singlet-octet and nonstrange-strange mixing angles:
\begin{equation}
\label{eq20}
\theta_P \,=\, -12.4^{\circ} \pm 3.5^{\circ},~~~
\varphi_P \,=\, 42.3^{\circ} \pm 3.5^{\circ},
\end{equation}
which are compared in Table~\ref{tab1} with the mixing angle values obtained 
from the well established and accepted phenomenology and from the 
experimental data available at present.
Quite remarkably, our values (\ref{eq20}) are compatible within the errors
with the values for $\theta_P$ and $\varphi_P$ obtained from very different 
physical processes (Table~\ref{tab1}).

Our values (\ref{eq17})-(\ref{eq18}) for the numbers $k(\phi)$ and 
$k(f_2^{\prime})$ correspond to the following values of the vector 
($\varphi_V$) and tensor ($\varphi_T$) nonstrange-strange mixing angles 
(our analysis is not sensitive to the sign of $\varphi$):
\begin{equation}
\label{eq21}
\vert\varphi_V\vert \,=\, 10^{\circ} \pm 8^{\circ},~~~
\vert\varphi_T\vert \,=\, 16^{\circ} \pm 11^{\circ},
\end{equation}
which are very close to the predictions 
$\vert\varphi_V\vert \simeq 3.4^{\circ}$ and 
$\vert\varphi_T\vert \simeq 7.3^{\circ}$
of most theoretical and phenomenological analyses (see, for example, 
Refs. \cite{R1,R2,R12,R13,R14,R15}) including the quadratic and linear GMO 
mass formulae \cite{PDG}. 
The values (\ref{eq21}) are also compatible with the ``ideal'' mixing angle
$\varphi^{ideal}=0$.

The value of the mixing angle $\varphi_S$ for the $f_0$(980) scalar meson 
is found from Eq. (\ref{eq12}) with the value (\ref{eq19}) for the 
number $k(f_0)$:
\begin{equation}
\label{eq22}
\vert\varphi_S\vert \,=\, 13^{\circ} \pm 9^{\circ}.
\end{equation}
This value of the mixing angle $\varphi_S$ is not consistent with the 
results of recent phenomenological studies. 
For example, the present data for decays of $f_0$(980) into the channels 
$\pi\pi$, $K\bar{K}$ and $\gamma\gamma$ allow to interpret the $f_0$(980) 
as either the quasi-singlet state with $\varphi_S \simeq  55^{\circ}$ or the 
quasi-octet one with $\varphi_S \simeq  -43^{\circ}$ \cite{I8}.
Also the studies of the decay processes $\phi \rightarrow \pi^0\pi^0\gamma$, 
$J/\psi \rightarrow \omega\pi\pi$ \cite{I11} and 
$D_s^+ \rightarrow f_0(980)\pi^+$ \cite{I9} have led 
(in terms of $\varphi_S$ defined by Eq.~(\ref{eq6})\,) 
to the values of $\varphi_S \simeq 70^{\circ}$ 
and $\varphi_S \simeq 76^{\circ}$, respectively. 

The obtained disagreement for the $f_0$(980) can probably be related with a 
question which is still open. 
This is whether the $f_0$(980) belongs to the scalar $q\bar{q}$ 
nonet $1{\,}^3P_0$ or whether it should be considered as an exotic state.
The scalar nonet classification can be performed in terms of so-called
``bare states'' (the $K$-matrix poles) corresponding to $q\bar{q}$ states 
``before'' the mixing which is caused by the transitions
$q\bar{q}~state \rightarrow real~mesons$ \cite{K1,K2,K3,I7}. 
In this way \cite{I7} the scalar $q\bar{q}$ nonet $1{\,}^3P_0$ has been found 
to be: $f_0^{bare}(720\pm100)$, $f_0^{bare}(1260\pm30)$, 
$a_0^{bare}(960\pm30)$ and $K_0^{bare}(1220_{-150}^{+~50})$.
The flavour wave functions of the orthogonal states
$f_0^{bare}(720)$ and $f_0^{bare}(1260)$ are given in the form
(\ref{eq6}) by the mixing angle 
$\varphi_S[f_0^{bare}(720)] = {-70^{\circ\,}}_{-16^{\circ}}^{+~5^{\circ}}$
\cite{I7}. The transitions $q\bar{q}~state \rightarrow real~mesons$ mix these 
bare states with each other as well as with nearly states. The real 
$f_0$(980) meson is a result of this mixing and is related both to
the $f_0^{bare}(720\pm100)$ and $f_0^{bare}(1260\pm30)$ states.

\begin{figure}[hbtp]
\centering\mbox{\epsfig{file=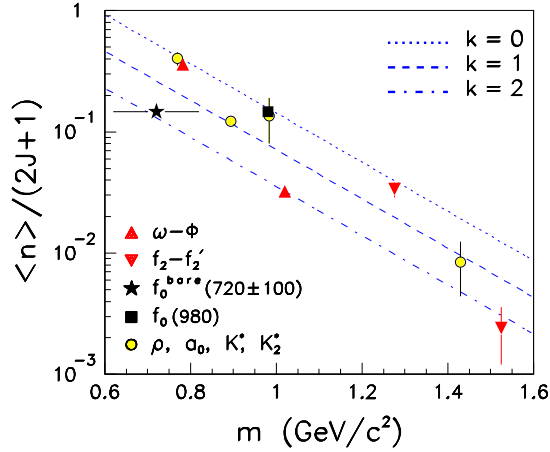,width=0.5\textwidth}}
\caption{\small Total production rate per spin and isospin state as a function 
of the mass for vector, tensor and scalar mesons in hadronic Z decays. Curves 
are the result of the fit~4 (see text).} 
\label{fig2}
\end{figure}

There is some interesting relation between our phenomenology and the 
$K$-matrix analysis which can be illustrated by Fig.~\ref{fig2},
where the total production rates per spin and isospin state for vector, 
tensor and scalar mesons are plotted as a function of $m$ and three curves 
with $k=0$, $k=1$ and $k=2$ are the result of our final fit (fit~4).
According to (\ref{eq22}) the data point for the $f_0$(980) is close to the 
curve with $k=0$.
As was observed by L.~Montanet \cite{MPC} using the figure from Ref. \cite{U3},
the $f_0$(980) data point shifted 
from the real mass ($m$ = 980 MeV/$c^2$) to the bare mass
($m^{bare}$ = $720\pm100$ MeV/$c^2$) is close to the curve with $k=2$ 
(Fig.~\ref{fig2}).
This fact can probably
be considered as a speculative argument to re-determine the strangeness 
content of the $f_0$(980) using our phenomenology, but replacing the real mass 
of the $f_0$(980) to the bare one. Therefore the fit~4 is repeated with
this replacement and gives the following value of the scalar mixing angle
$\varphi_S^{\,bare}$:
\begin{equation}
\label{eq23}
\vert\varphi_S^{\,bare}\vert \,=\, 73^{\circ} \pm 7^{\circ} \pm 24^{\circ},
\end{equation}
where the second error is due to the uncertainty $\pm100$ MeV/$c^2$ of the 
bare mass. This value is well consistent with the predictions of the $K$-matrix 
analysis \cite{K2,I7} and of the recent phenomenological studies 
\cite{I8,I9,I11}.
It can be noted also that there is only one more scalar meson in our analysis, 
i.e. $a_0$(980), but the real mass of this scalar is very close to the bare 
mass of the $a_0^{bare}(960\pm30)$ state.
So, if the $f_0$(980) is mostly $s\bar{s}$ state,
our results for the scalar mesons suggest that its total production rates 
are probably given by the bare masses.

In conclusion, the new phenomenological approach has been suggested to 
determine the strangeness contents of light-flavour isoscalars.
For the first time for this purpose the total production rates per hadronic 
Z decay of all light-flavour hadrons measured so far at LEP were used.
So, the new physical process (${\rm Z}^0 \rightarrow hadrons$) is added
to the list of ones used in the world available analyses on the 
mixing angle determination. 
Assuming the one-mixing-angle scheme, the orthogonality of the isoscalar 
partners and no mixing with other states and glueballs, our approach 
gives us the following values of the pseudoscalar, vector and tensor mixing 
angles: 
$\varphi_P = 42.3^{\circ} \pm 3.5^{\circ}$ 
($\theta_P = -12.4^{\circ} \pm 3.5^{\circ}$),
$\vert\varphi_V\vert = 10^{\circ} \pm 8^{\circ}$ and
$\vert\varphi_T\vert = 16^{\circ} \pm 11^{\circ}$,
in good agreement with the present experimental evidence. 
The same analysis gives us two values of the scalar mixing angle:
$\vert\varphi_S\vert = 13^{\circ} \pm 9^{\circ}$ 
if the $f_0$(980) is taken with the real mass, 
but $\vert\varphi_S^{\,bare}\vert = 73^{\circ} \pm 7^{\circ} \pm 24^{\circ}$
if the $f_0$(980) is taken with the bare mass ($K$-matrix pole) of the 
$f_0^{bare}(720\pm100)$ state.
Only the second value is consistent with recent phenomenological analyses. 
{\it If their conclusions are correct}, it means that in the framework
of our approach the total production rates of scalar mesons are probably
given by the bare masses.

\bigskip
\noindent
{\it Acknowledgments.} I am grateful to L.~Montanet for interesting and
inspiring discussions and for useful comments.


\begin{thebibliography}{99}
\bibitem{I1}
Th. Feldmann, 
Int.\,J.\,Mod.\,Phys. {\bf A\,15} (2000) 159.
\bibitem{I10}
L. Montanet, 
Nucl.\,Phys.\,(Proc.\,Suppl.) {\bf B\,86} (2000) 381.
\bibitem{R1}
A. Bramon, R. Escribano, M.D. Scadron, 
Eur.\,Phys.\,J. {\bf C\,7} (1999) 271.
\bibitem{I2}
F.J. Gilman, R. Kauffman, 
Phys.\,Rev. {\bf D\,36} (1987) 2761.
\bibitem{I5}
J.F. Donoghue, B.R. Holstein, Y.-C.R. Lin, 
Phys.\,Rev.\,Lett. {\bf 55} (1985) 2766.
\bibitem{I6}
J. Schechter, A. Subbaraman, H. Weigel, 
Phys.\,Rev. {\bf D\,48} (1993) 339.
\bibitem{R11}
P. Ball, J.-M. Fr\`ere, M. Tytgat,
Phys.\,Lett. {\bf B\,365} (1996) 367.
\bibitem{I3}
L. Burakovsky, T. Goldman, 
Phys.\,Lett. {\bf B\,427} (1998) 361.
\bibitem{I4}
F.G. Cao, A.I. Signal, 
Phys.\,Rev. {\bf D\,60} (1999) 114012.
\bibitem{I7}
V.V. Anisovich et al.,
Phys.\,Atom.\,Nucl. {\bf 63} (2000) 1410.
\bibitem{I8}
V.V. Anisovich, L. Montanet, V.N. Nikonov, 
Phys.\,Lett. {\bf B\,480} (2000) 19.
\bibitem{I9}
E. van Beveren, G. Rupp, M.D. Scadron, 
Phys.\,Lett. {\bf B\,495} (2000) 300.
\bibitem{R10}
Th. Feldmann, P. Kroll, B. Stech,
Phys.\,Rev. {\bf D\,58} (1998) 114006.
\bibitem{R2}
A. Bramon, R. Escribano, M.D. Scadron, 
Phys.\,Lett. {\bf B\,503} (2001) 271.
\bibitem{R3}
T.N. Pham, 
Phys.\,Lett. {\bf B\,246} (1990) 175.
\bibitem{R12}
M. Benayoun et al.,
Phys.\,Rev. {\bf D\,59} (1999) 114027.
\bibitem{R4}
Mark III Collab., D. Coffman et al., 
Phys.\,Rev. {\bf D\,38} (1988) 2695. 
\bibitem{R5}
DM2 Collab., J. Jousset et al., 
Phys.\,Rev. {\bf D\,41} (1990) 1389. 
\bibitem{R9}
V.V. Anisovich et al.,
Phys.\,Lett. {\bf B\,404} (1997) 166.
\bibitem{R6}
W.D. Apel et al.,
Phys.\,Lett. {\bf B\,83} (1979) 131.
\bibitem{R7}
N.R. Stanton et al.,
Phys.\,Lett. {\bf B\,92} (1980) 353.
\bibitem{R8}
Crystal Barrel Collab., C. Amsler et al.,
Phys.\,Lett. {\bf B\,294} (1992) 451.
\bibitem{PDG}
{\it Particle Data Group}, D.E. Groom et al.,
Eur.\,Phys.\,J. {\bf C\,15} (2000) 1.
\bibitem{R13}
H.F. Jones, M.D. Scadron,
Nucl.\,Phys. {\bf B\,155} (1979) 409.
\bibitem{R14}
G. Dillon, G. Morpurgo,
Z.\,Phys. {\bf C\,64} (1994) 467.
\bibitem{R15}
D.M. Li, H. Yu, Q.-X. Shen,
J.\,Phys. {\bf G\,27} (2001) 807.
\bibitem{I11}
R. Delbourgo, D. Liu, M.D. Scadron, 
Phys.\,Lett. {\bf B\,446} (1999) 332.
\bibitem{I12}
N.A. T\"ornqvist, M. Roos, 
Phys.\,Rev.\,Lett. {\bf 76} (1996) 1575.
\bibitem{I13}
S. Ishida et al., 
Prog.\,Theor.\,Phys. {\bf 98} (1997) 621.
%
\bibitem{U1}
V. Uvarov, 
Proc. of the 15th Int. Conf. on Particles and Nuclei (Uppsala, 1999),
Eds. G. F\"aldt et al., Nucl.\,Phys. {\bf A\,663} (2000) 633.
\bibitem{D9}
DELPHI Collab., P. Abreu et al., Phys.\,Lett. {\bf B\,475} (2000) 429.
\bibitem{U2}
V. Uvarov, 
Phys.\,Lett. {\bf B\,482} (2000) 10.
\bibitem{U3}
V. Uvarov, 
Proc. of the 35th Rencontre de Moriond on QCD and High Energy Hadronic
Interactions (Les Arcs, 2000); hep-ex/0011058.
\bibitem{U4}
V. Uvarov, 
Proc. of the 30th Int. Symp. on Multiparticle Dynamics (Tihany, 2000);
hep-ph/0011239; preprint IHEP 2000-47, Protvino, 2000.
%
\bibitem{A1}
ALEPH Collab., R. Barate et al., Phys.\,Rep. {\bf 294} (1998) 1.
\bibitem{A2}
ALEPH Collab., R. Barate et al., Eur.\,Phys.\,J. {\bf C\,5} (1998) 205.
\bibitem{A3}
ALEPH Collab., R. Barate et al., Eur.\,Phys.\,J. {\bf C\,16} (2000) 613.
%
\bibitem{D1}
DELPHI Collab., P. Abreu et al., Phys.\,Lett. {\bf B\,361} (1995) 207.
\bibitem{D2}
DELPHI Collab., P. Abreu et al., Z.\,Phys. {\bf C\,65} (1995) 587.
\bibitem{D3}
DELPHI Collab., P. Abreu et al., Z.\,Phys. {\bf C\,67} (1995) 543.
\bibitem{D4}
DELPHI Collab., W. Adam et al., Z.\,Phys. {\bf C\,69} (1996) 561.
\bibitem{D5}
DELPHI Collab., W. Adam et al., Z.\,Phys. {\bf C\,70} (1996) 371.
\bibitem{D6}
DELPHI Collab., P. Abreu et al., Z.\,Phys. {\bf C\,73} (1996) 61.
\bibitem{D7}
DELPHI Collab., P. Abreu et al., Eur.\,Phys.\,J. {\bf C\,5} (1998) 585.
\bibitem{D8}
DELPHI Collab., P. Abreu et al., Phys.\,Lett. {\bf B\,449} (1999) 364.
%
\bibitem{L1}
L3 Collab., M. Acciarri et al., Phys.\,Lett. {\bf B\,328} (1994) 223.
\bibitem{L2}
L3 Collab., M. Acciarri et al., Phys.\,Lett. {\bf B\,393} (1997) 465.
\bibitem{L3}
L3 Collab., M. Acciarri et al., Phys.\,Lett. {\bf B\,407} (1997) 389.
\bibitem{L4}
L3 Collab., M. Acciarri et al., Phys.\,Lett. {\bf B\,479} (2000) 79.
%
\bibitem{O1}
OPAL Collab., P.D. Acton et al., Phys.\,Lett. {\bf B\,305} (1993) 407.
\bibitem{O2}
OPAL Collab., R. Akers et al., Z.\,Phys. {\bf C\,63} (1994) 181.
\bibitem{O3}
OPAL Collab., G. Alexander et al., Phys.\,Lett. {\bf B\,358} (1995) 162.
\bibitem{O4}
OPAL Collab., R. Akers et al., Z.\,Phys. {\bf C\,67} (1995) 389.
\bibitem{O5}
OPAL Collab., R. Akers et al., Z.\,Phys. {\bf C\,68} (1995) 1.
\bibitem{O6}
OPAL Collab., G. Alexander et al., Z.\,Phys. {\bf C\,73} (1997) 569.
\bibitem{O7}
OPAL Collab., G. Alexander et al., Z.\,Phys. {\bf C\,73} (1997) 587.
\bibitem{O8}
OPAL Collab., K. Ackerstaff et al., Eur.\,Phys.\,J. {\bf C\,4} (1998) 19.
\bibitem{O9}
OPAL Collab., K. Ackerstaff et al., Eur.\,Phys.\,J. {\bf C\,5} (1998) 411.
%
\bibitem{K1}
V.V. Anisovich, Yu.D. Prokoshkin, A.V. Sarantsev,
Phys.\,Lett. {\bf B\,389} (1996) 388.
\bibitem{K2}
A.V. Anisovich, V.V. Anisovich, A.V. Sarantsev,
Z.\,Phys. {\bf A\,359} (1997) 173.
\bibitem{K3}
A.V. Anisovich, A.V. Sarantsev,
Phys.\,Lett. {\bf B\,413} (1997) 137.
\bibitem{MPC}
L. Montanet, private communication.
%
\end{thebibliography}
\end{document}